\documentclass[twocolumn,secnumarabic,amssymb,nobibnotes,aps,prs,showpacs,nofootinbib,superscriptaddress]{revtex4-1}
\usepackage[english]{babel}
\usepackage{graphicx}
\usepackage{amsmath}
\usepackage{amsfonts}
\usepackage[usenames,dvipsnames]{color}
\usepackage{hyperref}
\hypersetup{colorlinks}
\hypersetup{linkcolor=black, citecolor=black, urlcolor=blue}

\begin{document}
\title{Density-induced reentrant melting of colloidal Wigner crystals}
\author{J. C. Everts}
\email{j.c.everts@uu.nl}
\address{Institute for Theoretical Physics, Center for Extreme Matter and Emergent Phenomena,  Utrecht University, Leuvenlaan 4, 3584 CE Utrecht, The Netherlands}
\author{N. Boon}
\address{Division of Physical Chemistry, Lund University, Lund SE-221 00, Sweden}
\author{R. van Roij}
\address{Institute for Theoretical Physics, Center for Extreme Matter and Emergent Phenomena,  Utrecht University, Leuvenlaan 4, 3584 CE Utrecht, The Netherlands}
\date{\today}

\begin{abstract}
Electrostatic repulsions can drive crystallization in many-particle systems. For charged colloidal systems, the phase boundaries as well as crystal structure are highly tunable by experimental parameters such as salt concentration and pH. By using projections of the colloid-ion mixture to a system of (soft) repulsive spheres and the one-component plasma (OCP), we study the hitherto unexplained experimentally observed reentrant melting of electrostatically repelling colloids upon increasing the colloid density. Our study shows that the surface chemistry should involve a competition between adsorption of cations and anions to explain the observed density-induced reentrant melting.
\end{abstract}

\maketitle

\section{Introduction}
Crystalline ordering is observed in systems with building blocks as small as electrons \cite{Wigner:1934,Grimes:1979}, and as big as granular particles \cite{Torquato:2010}. The crystallization of a many-body system is a direct result of the forces between the particles, and relating these forces to the observed ordering in the system is a challenging problem in general. For example, crystallization can be driven by attractive interactions between the individual components, such as in the gas-crystal transition for a Lennard-Jones system below the triple-point temperature \cite{HansenMcDonald}. A purely repulsive interaction can also result in crystalline order, a text-book example being the self-assembly of colloidal hard spheres into a face-centered cubic crystal at high densities \cite{Alder:1957,Jones:1964, Pusey:1986}. In systems of charged particles, where Coulomb interactions are pivotal for understanding the crystallization transition, the inherent structure may also either form due to the attractions between oppositely charged species \cite{Leunissen:2005}, or due to mutual repulsion between like-charged particles. Crystallization due to electrostatic repulsions has been studied extensively for colloidal systems \cite{Williams:1974, Sirota:1989, Monovoukas:1989, Okubo:1990, Okubo:1991, Palberg:1994, Anand:2007, Herlach:2010}. These colloidal Wigner crystals \cite{Lindsay:1982} are very interesting from a engineering perspective as they can have extraordinary optical \cite{Kim:2011} or mechanical properties \cite{You:2009}. Due to the relatively large size of the particles, the transition to a crystal is easy to study in the lab. Interestingly, the charge of these particles is usually not fixed, but regulated \cite{Palberg:2006, Royall:2006, Dufresne:2009}, as it results from the chemical equilibrium between the ionizable surface and the ions in the solvent \cite{Ninham:1971}, which causes the effective forces between the particles to be highly tunable by experimentally controllable parameters. The ordering of the particles therefore shows strong dependence on parameters such as pH and salt concentration. This opens up a vast parameter space in which various crystalline structures can be found.

In earlier studies the formation of charge-induced crystallization was described within Poisson-Boltzmann theory \cite{Andelman} in the spherical- cell approximation. The resulting electrostatic potential and ionic diffuse screening layer around the central particle are then mapped through a renormalized colloidal charge to effective one-component systems for which the phase diagram is known empirically from e.g. simulations of point-Yukawa particles \cite{Smallenburg:2011}. Recently, it was shown that the cell approach can also be used to define a mapping to a one-component plasma (OCP) \cite{Boon:2015}, such that the freezing criterion for the classical OCP can be applied. Combining explicit models for the surface chemistry of the colloidal particles with the OCP mapping yields a model that fits very well with experiments and provides insight not only at the level of molecular details of the charging mechanism and the equilibrium constant, but also at the macroscopic scale of phase diagrams \cite{Boon:2015}. In this work we build on Ref. \cite{Boon:2015} by considering how various crystallization mechanisms affect the phase diagrams of colloidal particles that are subject to charge regulation. We will also connect our theoretical results with the experimental phase diagrams of Refs. \cite{Royall:2003, Royall:2006, Anda:2015}. We will highlight the well-understood reentrant melting as function of salt concentration, and the less understood reentrant melting as function of colloid density. We will show here that the latter can only be explained by a sufficiently strong density dependence of the colloidal charge and the screening length \cite{Royall:2006, Anda:2015}, that results from a binary adsorption model.

\begin{figure}[t]
\includegraphics[width=0.5\textwidth]{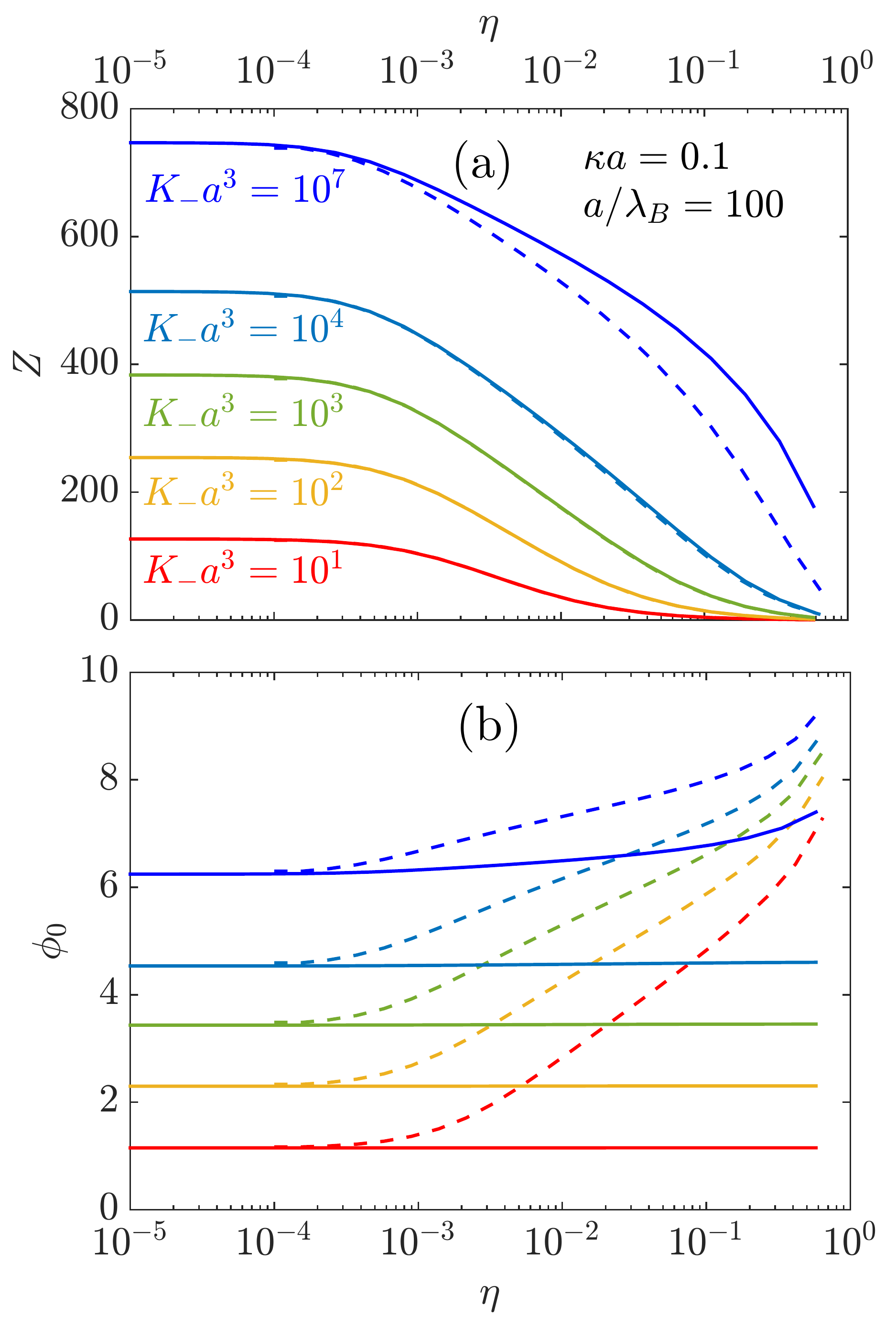}
\caption{The full curves represent (a)the particle charge $Z$ and (b) the dimensionless surface potential $\phi_0$ as function of packing fraction $\eta$ for various values of the equilibrium constant $K_-$, while keeping $K_+a^3=1$ and the total number of surface sites $M=10^7$ fixed. The dashed lines in (a) are the result of a constant-potential system where we chose $\phi_0$ such that the resulting $Z$ coincides in the dilute limit. The reverse is done in (b) but then for a constant-charge system that has a $Z$ such that the resulting $\phi_0$ coincides in the dilute limit. The lines for $K_-a^3=10^7$ are essentially the same as for the limit $K_-a^3\rightarrow\infty$, in which no anions can adsorb.}
\label{fig:charge}
\end{figure}

\section{Model}
Our description of the colloidal suspension invokes the spherical cell approximation as was introduced by Alexander \emph{et al.} in 1984  \cite{Alexander:1984}. In this approximation the suspension is divided into spherical cells, each containing one colloidal particle of radius $a$. These cells are all identical with radius $R$, which is related to the colloidal packing fraction via $\eta=(a/R)^3$. Within mean-field theory the ion-density profiles $\rho_\pm(r)$ are related to the electrostatic potential $\phi(r)/(\beta e)$ by $\rho_\pm(r)=\rho_s\exp[\mp\phi(r)]$, with $r$ the radial coordinate, $e$ the proton charge, $\beta^{-1}=k_BT$ the thermal energy and $\rho_s$ the salt concentration of a reservoir with which the system is assumed to be in osmotic equilibrium. Notice that varying $\rho_s$ is equivalent to varying the chemical potential of the ions. Apart from the particle radius $a$, there are two length scales in our problem. These are the Bjerrum length $\lambda_B=\beta e^2/4\pi\epsilon_\text{vac}\epsilon$ and the Debye screening length $\kappa^{-1}=(8\pi\lambda_B\rho_s)^{-1/2}$. Here $\epsilon_\text{vac}$ is the vacuum dielectric constant and $\epsilon$ is the relative dielectric constant in the solvent. Combining the mean-field density profiles with the Poisson equation results in the non-linear spherically symmetric Poisson-Boltzmann (PB) equation
\begin{equation}
\frac{d^2\phi}{dr^2}+\frac{2}{r}\frac{d \phi}{dr}=\kappa^2\sinh[\phi(r)], \quad r\in[a,R], 
\end{equation}
with boundary conditions $\phi'(a)=-Z\lambda_B/a^2$ and $\phi'(R)=0$, where a prime denotes a radial derivative and $Ze$ is the colloidal charge. From the solution of the PB equation the surface potential $\phi_0=\phi(a)$ and the Donnan potential $\phi_D=\phi(R)$ are found self-consistently once $Z$ is known. Here we do not only consider the constant-charge case where $Z$ is a known input parameter, but we will also calculate $Z$ self-consistently for charge regulation cases where we consider an associative charging mechanism in which a single surface site S can be occupied by negative ions N$^-$ and positive ions P$^+$. These are governed by the reactions $\text{S}+\text{P}^+ \leftrightarrows \text{SP}^+$ with equilibrium constant $K_+=[\text{S}][\text{P}^+]/[\text{SP}^+]$ and $\text{S}+\text{N}^- \leftrightarrows \text{SN}^-$ with equilibrium constant $K_-=[\text{S}][\text{N}^-]/[\text{SN}^-]$. This results in an adsorption isotherm that relates the surface charge to the surface potential \cite{Ninham:1971} via
\begin{equation}
\frac{Z}{M}=\frac{K_-\exp(-\phi_0)-K_+\exp(\phi_0)}{\sum_\sigma K_\sigma\exp(z_\sigma\phi_0)+K_+ K_-/\rho_s},
\label{eq:isotherm}
\end{equation}
with $z_\pm=\pm 1$, which reduces to the familiar Langmuir form
\begin{equation}
\frac{Z}{M}=\frac{1}{1+K_+/\rho_s\exp(\phi_0)},
\label{eq:posadsorb}
\end{equation}
in the limit where $K_-\rightarrow\infty$. Here $M$ is the number of sites available for adsorption and we will set it to $M=10^7$, which is equivalent to roughly one surface group per 1 nm$^2$ for a micron-sized particle. Note that high values for $K_\pm/\rho_s$ yield little tendency for the ions to adsorb, while small values for $K_\pm/\rho_s$ results in a significant fraction of occupied surface groups.  In Figure \ref{fig:charge} we plot some results for (a) $Z$ and (b) $\phi_0$ obtained from the cell model as function of $\eta$ for $M=10^7$, $a/\lambda_B=100$ and $\kappa a =0.1$. We see that the colloidal particle always discharges as function of $\eta$, while the corresponding $\phi_0$ increases or is approximately constant for the case in which both positive as well as negative ions can adsorb. The observed constant potential in the latter case is because both the chargeabilities $Y_\pm=\kappa aM/(8\pi K_\pm a^3)$ are much larger than unity for these parameters, as we shall show in the Appendix. 

From the cell model it is possible to extract effective pair potentials, which we will use in various freezing criteria. The best known route towards an effective pair potential for (highly) charged particles uses charge renormalization \cite{Trizac:2002, Diehl:2005} in combination with DLVO theory \cite{Derjaguin:1948, VerweyOverbeek}
. An effective charge $Z^*$ is defined by extrapolating the linear screening solution fitted to the numerical solution for the far-field electrostatic potential to $r=a$ \cite{Alexander:1984}, 
\begin{align}
Z^*=&\frac{\tanh\phi_D}{\bar\kappa \lambda_B}\big[(\bar\kappa^2 a R-1)\sinh(\bar\kappa R-\bar\kappa a) \nonumber 
\\ &+(\bar{\kappa}R-\bar{\kappa}a)\cosh(\bar\kappa R-\bar\kappa a)\big]. \label{eq:Zren}
\end{align}
Here $\bar\kappa^{-1}$ is a colloid-density dependent screening length given by $\bar{\kappa}^2=\kappa^2\cosh\phi_D$. From this so-called renormalized charge $Z^*$, we can define the effective DLVO pair potential 
\begin{equation}
\beta U_\text{DLVO}(r)=\begin{cases}
\infty, \quad r<2a, \\
\left(\dfrac{Z^*\exp(\bar{\kappa} a)}{1+\bar{\kappa} a}\right)^2\dfrac{\lambda_B\exp(-\bar{\kappa} r)}{r}, \quad r\geq2a. \end{cases} \label{eq:DLVO}
\end{equation}
We remark here that this DLVO-based method is known to become inaccurate for dense systems \cite{Loewen:1992,vanRoij3:1997,Warren:2000,Hansen:2000,Gruenberg:2001, BoonPNAS:2015}, due to many-body effects resulting from a significant overlap between double layers with the hard core of other particles. Nevertheless, in case the double layers constitute a relatively thin shell around the individual particles, the DLVO form is deemed accurate.

Alternatively, one can choose to calculate effective point-Yukawa charges by fitting the linearized solution for the ``far-field" electrostatic potential in the cell to the non-linear solution that follows from the Poisson-Boltzmann model. By extrapolating the linear solution to $r=0$ , an effective Yukawa point charge $Q$ can be identified in the origin \cite{BoonPNAS:2015}, which is found to depend on the cell-boundary parameters via
\begin{equation}
Q = \dfrac{\tanh\phi_D}{\bar{\kappa} \lambda_{B}} \left[\bar\kappa R \cosh (\bar \kappa R) - \sinh (\bar\kappa R)\right] .\label{eq:yukawacharge}
\end{equation}
Using this effective point charge, the effective pair interaction can be expressed as the sum of the non-electrostatic hard-core repulsion and a Yukawa potential
\begin{equation}
\beta U_\text{Y}(r)=\begin{cases}
\infty, \quad r<2a, \\
\dfrac{Q^2 \lambda_B \exp(-\bar\kappa r)}{r}
, \quad r\geq2a. \end{cases} \label{eq:pointyuk}
\end{equation}
By means of computer simulations of these point particles \cite{BoonPNAS:2015} it has been confirmed recently that this approach yields a very accurate estimate for the colloid-colloid pair-correlation functions and the pressure for both dilute and dense colloidal systems when compared with a mixture of colloids and ions.

\section{Crystallization criteria}
There are various approaches towards determining the location of the charge-induced crystallization transition that we will discuss in this work. As it has been proposed earlier by other authors, it is tempting to view the charged system as an effective hard-sphere system, with an effective hard-core diameter $\sigma_\text{eff}$ that is larger than the original diameter of the particle due to the electrostatic repulsions \cite{Stigter:1954, vanMegen:1975,Brenner:1976,Beunen:1981}. This can, for example, be achieved by defining the second virial coefficient $B_2=(1/2)\int d{\bf r}\{1-\exp[-\beta U_\text{DLVO}({\bf r})]\}$ and imposing the second virial coefficient of the hard-sphere fluid $B_2=(2/3)\pi\sigma_\text{eff}^3$. By using that the hard-sphere system crystallizes at packing fraction $(\pi/6)\sigma_\text{eff}^3\rho>0.5$ \cite{Pusey:1986}, we arrive at the freezing criterion $B_2\rho>2$, with $\rho$ the colloid density. 

Another approach that does not rely on any pair potential can be found by comparing the osmotic pressure $\Pi$ to that of a system of hard spheres. Within the cell model, the osmotic pressure is given by summing the ionic and the hard-core contributions. This results in  $\beta\Pi/\rho=2\rho_s(\cosh\phi_D-1)/\rho+(1+\eta+\eta^2-\eta^3)/(1-\eta)^3$, where we have used the Carnahan-Starling expression for the second term. The criterion $\beta\Pi/\rho>13$ can now be applied in analogy to the hard-sphere system. 

Both hard-sphere like criteria that are described above are expected to be accurate at high salt concentration, where the double layers and hence the repulsive interactions are short-ranged. On the other hand, when the interactions are longer ranged the effective hard-core model is expected to break down and other approaches are needed. Recent work describes a method to map the suspension to a system of point-Coulomb particles in a neutralizing background. The latter system is known as the one-component plasma (OCP). The mapping to an OCP constitutes a partitioning of the full ionic charge into individual double layers that (partially) neutralize the charged particles and a homogeneous background of ionic charge that neutralizes the remaining charge. Within the cell model, this background is identified as the ionic charge density on the cell boundary, $\rho_+(R)-\rho_-(R)$. It defines an equivalent OCP-point charge $Z_\text{OCP}$ via the charge-neutrality requirement, i.e., $Z_\text{OCP}=-[\rho_+(R)-\rho_-(R)]/\rho$, such that the OCP coupling parameter $\Gamma_\text{OCP}=Z_\text{OCP}^2\lambda_B\rho^{1/3}$, which is the dimensionless parameter that fully characterizes the OCP, takes the form
\begin{equation}
\Gamma_\text{OCP} = \frac{1}{16\pi^2} \frac{\tanh^2 \phi_D}{\bar{\kappa} \lambda_{B}} (\bar \kappa D)^5, \label{eq:OCP}
\end{equation}
where we used the mean interparticle distance $D=\rho^{-1/3}$, given within the cell model by $D^3 = 4 \pi R^3  /3$. For $\Gamma_\text{OCP}<106$ the OCP is in the disordered fluid state, yet for $\Gamma_\text{OCP}>106$ it favours a body-centered cubic (BCC) crystalline state \cite{Slattery:1980, Ichimaru:1982}. The empirical criterion $\Gamma_\text{OCP}>106$ was very recently shown to be very successful in describing experiments on colloidal systems \cite{Boon:2015} and is an attractive option due to its simplicity. 

\begin{figure*}[ht]
\includegraphics[width=\textwidth]{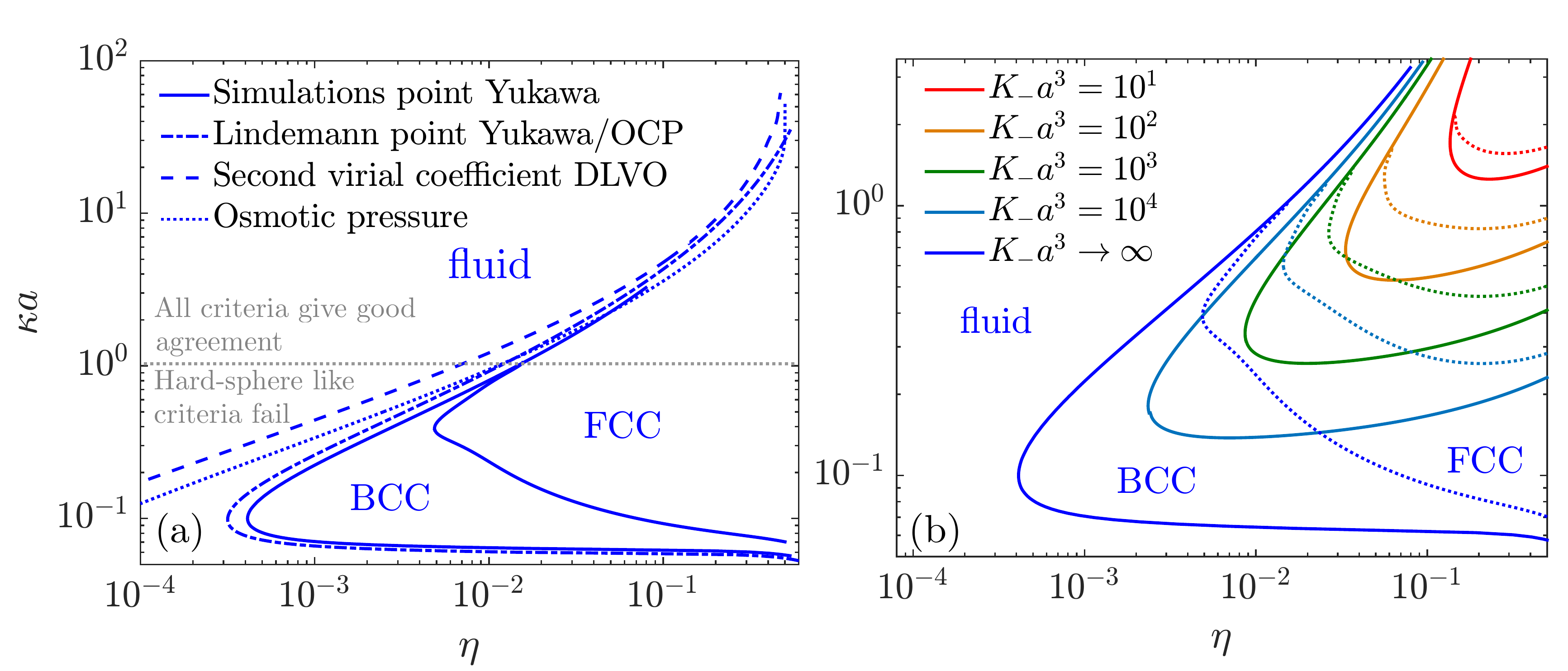}
\caption{(a) Phase diagram in the ($\eta,\kappa a$) representation for particles that acquire their charge through cationic adsorption only. We fix the Bjerrum length $\lambda_B=0.01a$ and equilibrium constant $K_+a^3=1$. The full lines stem from the extrapolated point charge $Q$ of the cell model, which was used as the point-Yukawa charge in the computer simulation criterion from Ref. \cite{Hynninen:2003}.  For this criterion, it is possible to obtain the crystal-crystal transition line from a FCC to BCC lattice. The dashed-dotted line is obtained by applying the Lindemann criterion for point-Yukawas ($\Gamma_\text Y=106$); this line overlaps with the result of a OCP criterion ($\Gamma_\text{OCP}=106$). The dashed line uses a second virial coefficient result ($B_2\rho=2$). Finally the dotted line uses an osmotic pressure criterion ($\beta\Pi/\rho=13$). A dotted grey line indicates the region below which the hard-sphere like criteria using $B_2$ and $\Pi$ are expected to fail. (b) Same as full lines in (a), but now we allow negative ions to adsorb, tuned by the equilibrium constant $K_-$. Notice that the full blue line is the same as in (a), yet here we use dotted lines for the FCC-BCC transitions for clarity. We use the same color coding as in Fig. \ref{fig:charge}.}
\label{fig:phase}
\end{figure*}

The OCP, however, does not feature a face-centered cubic (FCC) phase, such that it cannot capture the experimentally observed BCC-FCC phase transition. The effective Yukawa point charges defined by Eq. \eqref{eq:yukawacharge} and Eq. \eqref{eq:pointyuk} form an alternative approach to calculating the freezing lines in a charged colloidal system. We can use the Lindemann criterion for the effective pair potential \cite{Khrapak:2000} and find that crystalline order is expected if the so-called Yukawa coupling parameter
\begin{equation}
\Gamma_{\mathrm{Y}} \equiv \beta U_{\mathrm{Y}}(D) \left[1 + \bar \kappa D + \frac{1}{2}(\bar\kappa D)^2\right] \label{eq:yukawagamma}
\end{equation}
exceeds 106. Eq.~(\ref{eq:yukawagamma}), together with Eq. (\ref{eq:yukawacharge}) can be expanded in powers of $\kappa D$ to find
\begin{equation}
\Gamma_\mathrm{Y} = \frac{1}{16\pi^2} \frac{\tanh^2 \phi_D}{\bar{\kappa} \lambda_{B}} \left[(\bar \kappa D)^5 + \mathcal{O} ((\bar \kappa D)^7)\right].
\label{eq:c2c}
\end{equation}
Interestingly, up to 5th order in $\bar{\kappa}D$ this is just the coupling parameter $\Gamma_\text{OCP}$ from Eq. \eqref{eq:OCP} obtained by mapping the cell model to the one component plasma (OCP).  Point-Yukawa particles, however, do exhibit an FCC phase at sufficiently large $\bar{\kappa}D$, so it is interesting that the fluid-crystal lines from the OCP and point-Yukawa criteria coincide within numerical accuracy. Computer simulations of Yukawa systems \cite{Hamaguchi:1997,Hynninen:2003} have shown that the fluid-crystal transition (either to FCC or BCC) is accurately described by the condition
\begin{eqnarray}
\log [\beta U_{\mathrm{Y}}(D)] &=&  4.670 - 1.0417 \bar \kappa D + 0.1329 (\bar \kappa D)^2\nonumber\\
&-& 0.01043 (\bar \kappa D)^3 + 0.0004343 (\bar \kappa D)^4\nonumber\\
&-& 0.000006924 (\bar \kappa D)^5, \label{eq:pekkaa}
\end{eqnarray}
for $0<\bar\kappa D<12$, which is up to minor deviations equivalent to the Lindemann criterion of Eq. \eqref{eq:yukawagamma}. The criterion for the transition between a BCC to an FCC phase was found to be
\begin{eqnarray}
\log [\beta U_{\mathrm{Y}}(D)] &=&  97.65 - 151.469499 \bar \kappa D + 106.626405 (\bar \kappa D)^2\nonumber\\
&-& 41.67136 (\bar \kappa D)^3 + 9.639931 (\bar \kappa D)^4\nonumber\\
&-& 1.3150249 (\bar \kappa D)^5  + 0.09784811 (\bar \kappa D)^6\nonumber\\\  
&-& 0.00306396 (\bar \kappa D)^7,
\label{eq:pekka}
\end{eqnarray}
for $1.85 < \bar \kappa D <6.8$.
We remark that instead of the point-Yukawa approach we could also have used $U_\text{DLVO}(r)$ as was done in Ref. \cite{Smallenburg:2011}. Recent work, however, has shown that DLVO-based approaches underestimate the effective repulsion at high packing fractions, even if combined with methods such as charge renormalization. Indeed, $U_\text{DLVO}(r)$ was not able to accurately describe the experiments in Ref. \cite{Boon:2015}. The point-Yukawa approach therefore yields a more direct and accurate route to the effective screened-Coulomb interactions at any density. 

\begin{figure*}[t]
\includegraphics[width=\textwidth]{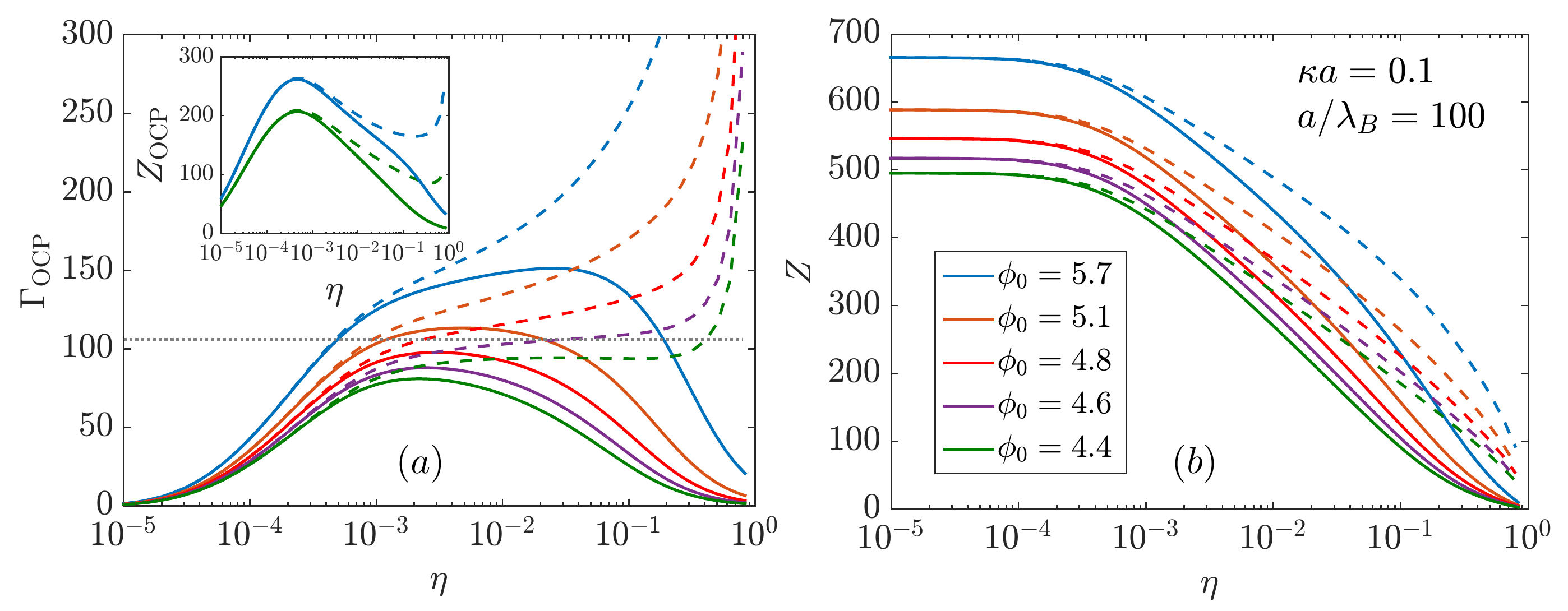}
\caption{(a) The OCP coupling parameter $\Gamma_\text{OCP}$ (see Eq. \eqref{eq:OCP}) as function of packing fraction $\eta$ for a charging mechanism $\text{S}+\text{P}^+ \leftrightarrows \text{SP}^+$ with equilibrium constant $K_+=[\text{S}][\text{P}^+]/[\text{SP}^+]$ (dashed lines), compared to a constant-potential system (full lines) with a chosen surface potential such that $Z$ (shown in (b)) coincides in the dilute limit. Observe that only for constant-potential particles $\Gamma_\text{OCP}$ can intersect the dotted grey line $\Gamma_\text{OCP}=106$ twice, showing that these systems exhibit reentrant melting as function of $\eta$. This is because constant-potential particles have the strongest tendency to decrease $Z_\text{OCP}$ for $\eta\gtrsim 10^{-3}$, as can be seen in the inset of (a) where $Z_\text{OCP}$ as function of $\eta$ is shown for the highest and lowest surface potential. This decrease in $Z_\text{OCP}$ occurs because constant-potential particles have a larger tendency to discharge, which is shown in (b) where we plot $Z$ as function of $\eta$.}
\label{fig:CPvsCR}
\end{figure*}

\section{Phase diagrams}
To get a better idea of the reliability of the various crystallization criteria for colloidal particles that are subjected to charge regulation, we will compare them first for the case when only positive ions can adsorb ($K_-\rightarrow\infty$). In Fig. \ref{fig:phase}(a), we plot a few phase boundaries in the $(\eta,\kappa a)$ representation for a variety of crystallization criteria for $a/\lambda_B=100$ and $K_+a^3=1$, featuring FCC, BCC and fluid phases. Note that only $\eta<0.5$ is shown. For these system parameters, we see that the BCC only appears in a finite ``pocket'' of in intermediate packing fractions $10^{-3}\lesssim\eta\lesssim10^{-1}$ and $10^{-1}\lesssim\kappa a\lesssim1$. The fluid-BCC line and the fluid-FCC line at $\kappa a\gtrsim 1$, as predicted by the Lindemann criterion Eq. \eqref{eq:yukawagamma}, are very close to the slightly more accurate simulation-based criterion of Eq. \eqref{eq:pekkaa} for all $\kappa a$. Notice furthermore that the FCC-BCC line is connected only to the fluid-crystal line of Eq. \eqref{eq:pekkaa}. Finally, as was mentioned in the previous section, the Lindemann criterion Eq. \eqref{eq:yukawagamma} is indeed equivalent to the OCP criterion of Eq. \eqref{eq:OCP}: we found that they essentially overlap within the numerical accuracy, and hence we have drawn them as a single line. 

The good predictive power of the OCP-like criteria of Eqs. \eqref{eq:OCP} and \eqref{eq:yukawagamma} does not come as a surprise for the regime of $\kappa a <1$, where the screening length is large compared to the particle size. However, the good quality of the OCP criterion  of Eq. \eqref{eq:OCP} at $\kappa a >1$ and even at $\kappa a>10$ is quite striking. Likewise, it should not come as a surprise that the hard-sphere freezing criteria based on $B_2\rho>2$ of the DLVO potential and the osmotic pressure $\Pi>13k_B T\rho$ of the cell model perform well at $\kappa a>10$ and reasonably well at $\kappa a \approx 1$. However, at $\kappa a<1$ they deviate substantially and are, therefore, not capable of predicting the empirical OCP-type criteria in this weak screening regime. In particular, the hard-sphere like criteria cannot capture the``back-bending" of the crystallization line to high $\eta$ at $\kappa a\lesssim 10^{-1}$, which is caused by the discharging of the particles at low salt concentrations, such that repulsions weaken and melting occurs \cite{Smallenburg:2011, Boon:2015}. The hard-sphere like criteria actually do show this bending-back phenomenon far below the scale of Fig. \ref{fig:phase}(a). The ``up-bending" of the crystallization line to high $\eta$ at $\kappa a\gtrsim 10^{-1}$ is due to the reduced repulsions which comes from the enhanced screening of the (increasing) colloidal charge. The resulting shape of the crystallization line of Fig. \ref{fig:phase}(a) describes a fluid-BCC-fluid or fluid-BCC-FCC-fluid phase sequence upon increasing $\kappa a$ at fixed $\eta\in(10^{-3},10^{-1}$), a reentrant melting that was also found in the constant-potential calculations of Ref. \cite{Smallenburg:2011} and in the experiments and calculations of Ref. \cite{Boon:2015}. However, particles described by the adsorption isotherm Eq. \eqref{eq:posadsorb} cannot account for the reentrant melting that was observed in Ref. \cite{Royall:2006} upon increasing the colloid density. Below we will show that an extension of the existing theories to include adsorption of a second ionic species does give rise to such a density-induced reentrant melting phenomenon.

On the basis of the superior performance of the OCP-based rather than the hard-sphere based crystallization criteria, we will now only consider the criteria of Eqs. \eqref{eq:pekkaa} and \eqref{eq:pekka}, which includes input from Eq. \eqref{eq:pointyuk}. We focus on the effect of anionic and cationic adsorption by setting $K_+a^3=1$ as before, together with setting a finite equilibrium constant $K_-$ (rather than the $K_-\rightarrow\infty$ limit which prevents anionic adsorption).

In Fig. \ref{fig:phase}(b) we show a set of phase diagrams, again in the $(\eta,\kappa a)$-representation, for a variety of $K_-$, showing fluid, BCC, and FCC states as expected. However, the crystallization lines at finite $K_-$ all exhibit a regime of $\kappa a$ where the phase sequence fluid-BCC-fluid appears upon increasing $\eta$. This density-induced reentrant melting is absent in the line for $K_-\rightarrow\infty$, which is the lowest-lying curve in Fig. \ref{fig:phase}(b). In other words, the feature of a reentrant fluid with increasing colloid concentration depends crucially on the existence of multiple charging mechanisms. Finally, we remark that for results with a non-zero $K_-^{-1}$, the crystallization boundary at $\eta=0.5$ is not correctly predicted because point particles do not exhibit hard-sphere crystallization.

We now try to rationalize the occurrence of a reentrant fluid as function of $\rho$. For this we investigate the OCP criterion, since it approximates the freezing lines of the Yukawa result (Eq. \eqref{eq:pekkaa}) accurately and it has the added advantage of providing a physical mechanism. In order to get a reentrant fluid, $\Gamma_\text{OCP}$ must be non-monotonous as function of density. For this we calculate
\begin{equation}
\frac{\partial\Gamma_\text{OCP}}{\partial\rho}=\Gamma_\text{OCP}\left(\frac{1}{3}Z_\text{OCP}\rho^{-1}+2\frac{\partial Z_\text{OCP}}{\partial\rho}\right),
\label{eq:gammadiff}
\end{equation}
and investigate its sign. The first term in Eq. \eqref{eq:gammadiff} is always positive. In the dilute limit the first term dominates in Eq. \eqref{eq:gammadiff}, hence $\Gamma_\text{OCP}$ increases with $\rho$. Compressing the system tends to reduce the mutual repulsions, the particles discharge and this reduces $Z_\text{OCP}$. This effect is not strong enough to drive $\partial_\rho \Gamma_\text{OCP}<0$ for particles that acquire their charge through adsorption of only a single ion species. However, $Z_\text{OCP}$ has a much stronger tendency to decrease for $\eta\gtrsim 10^{-3}$ if the particles have significant adsorption affinities for \emph{both} cations and anions, and therefore have a fairly constant surface potential. This is illustrated in Fig. \ref{fig:CPvsCR}(a), where $\Gamma_\text{OCP}$ is shown (with $Z_\text{OCP}$ in the inset) as function of $\eta$ for particles on which only positive ions can adsorb, and we compare these quantities with constant-potential particles. Indeed, we see that $\Gamma_\text{OCP}$ can intersect the line $\Gamma_\text{OCP}=106$ twice as function of $\eta$ for constant-potential particles. This can be rationalized from the fact that these type of particles have a larger tendency to discharge for $\eta\gtrsim 10^{-3}$ as is shown in Fig. \ref{fig:CPvsCR}(b), which reveals $Z$ as function of $\eta$ for constant-potential particles and for particles that acquire their charge by cationic adsorption. We conclude that for constant-potential particles the second term in Eq. \eqref{eq:gammadiff} can become sufficiently negative for a reentrant fluid to occur.

\section{Comparison with experiments}

\begin{figure}[h]
\includegraphics[width=0.5\textwidth]{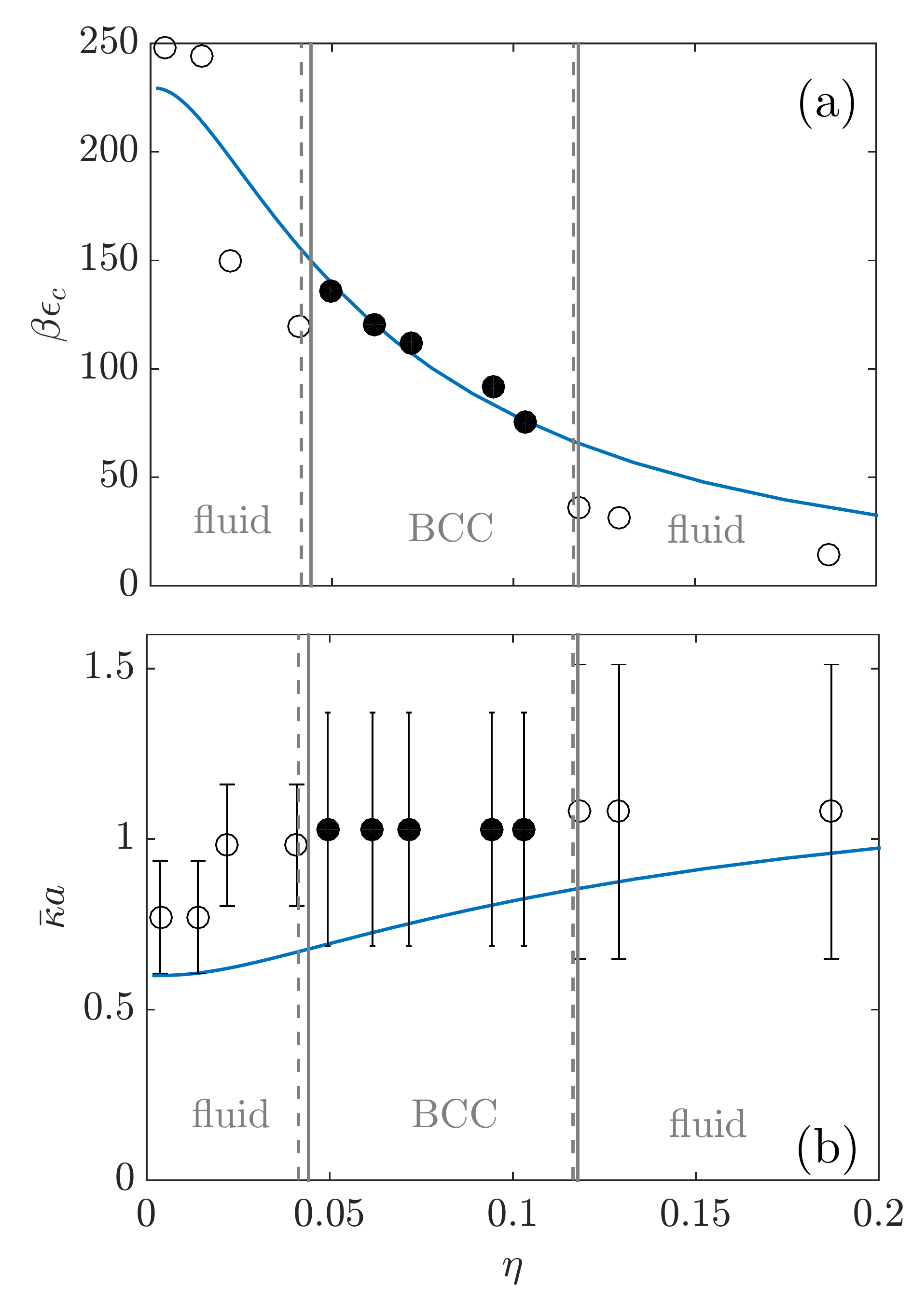}
\caption{Comparison of the phase boundaries  (fluid-BCC-fluid) and Yukawa parameters from Ref. \cite{Royall:2006}. The full grey lines are obtained from a cell-model calculation using $\kappa a=0.6$, $a/\lambda_B=125$, $K_-/K_+=51$ (equivalently $\phi_0=1.96$), while the dashed grey lines are from experiments. The equivalent contact value $\epsilon_c$ of the pair interaction potential from the cell model is shown as the full blue line in (a) and we compare them with Monte-Carlo simulations of  Ref. \cite{Royall:2006} using a DLVO potential, shown as the symbols. Open symbols correspond to fluid state points, while filled symbols are BCC state points. A similar comparison is made in (b) for $\bar{\kappa}a$, with error bars calculated from the data provided in Ref. \cite{Royall:2006}. We remark that the values of $\epsilon_c$ and $\bar{\kappa}$ for the BCC state points in  Ref. \cite{Royall:2006} were estimated using the Yukawa phase diagram and not directly determined from simulation.
}
\label{fig:mirjam}
\end{figure}

To verify our approach towards reentrant melting in suspensions of charged colloids, we compare our results with experiments on poly(methyl methacrylate) spheres $(a=1 \ \mu\text{m})$ in a solvent mixture of $20\%$ cis-decaline and $80\%$ cyclohexylbromide where a hitherto unexplained reentrant fluid was observed upon varying $\rho$, see Ref. \cite{Royall:2006,Anda:2015}. We use a Bjerrum length that is close to the experimental value $a/\lambda_B=125$ and vary the values of $\kappa a$ and $K_\pm a^3$ until good agreement with the experimental phase boundaries was obtained. It should be noted that in the parameter regime where we found a good fit, the last term in the denominator of Eq. \eqref{eq:isotherm} is small compared to the other terms in the denominator. This effectively means that we can only determine the ratio of $K_+/K_-$ rather than their individual values. Moreover, notice that Eq. \eqref{eq:isotherm} is independent of $\rho_s$ in this limit if $\phi_0$ is taken as an input parameter. This means that the calculated fluid-crystal boundary can also be explained by a constant-potential system for all salt concentrations.

The experimentally obtained fluid-BCC phase boundary at $\eta=0.0415$ and that of the BCC-reentrant fluid phase at $\eta=0.1165$ are represented by the dashed vertical lines in Fig. \ref{fig:mirjam}(a) and (b). The full vertical lines represent the corresponding phase boundary as predicted from our theory, using $\kappa a=0.6$ and $K_-/K_+=51$ as fit parameters, for which the dimensionless zeta potential reads $\phi_0=1.96$. Given that our theory is capable of predicting a reentrant fluid phase, it should not come as a surprise that we can fit the two experimentally observed phase boundaries in terms of these two fit parameters rather accurately.

Interestingly, however, the structure of various state points in both fluid phases and the BCC phase was also investigated in Ref. \cite{Royall:2006} by means of simulations of a system with a pairwise DLVO potential of the form of Eq. \eqref{eq:DLVO}. The contact potential $U_\text{DLVO}(2a)\equiv \epsilon_c$ and the effective screening length $\bar{\kappa}^{-1}$ were obtained from fits to the experimentally observed radial distribution function, and are represented by the open symbols in Fig. \ref{fig:mirjam}(a) and (b), respectively, where the error bars stem from Ref. \cite{Royall:2006}. The parameters $\epsilon_c$ and $\bar{\kappa}a$ for the crystal are shown as filled symbols, and were obtained from estimations by using the Yukawa phase diagram. The full curves in Fig. \ref{fig:mirjam} (a) and (b) represent our prediction of $\bar{\kappa}^{-1}$ and $\epsilon_c$, given within our calculation by $\epsilon_c=U_Y(2a)$, with the fit parameters obtained from the phase boundaries as discussed above. The agreement is very satisfactory and is an indication that the underlying charging mechanism indeed involves a competing cation and anion process.

\begin{figure}[t]
\includegraphics[width=0.5\textwidth]{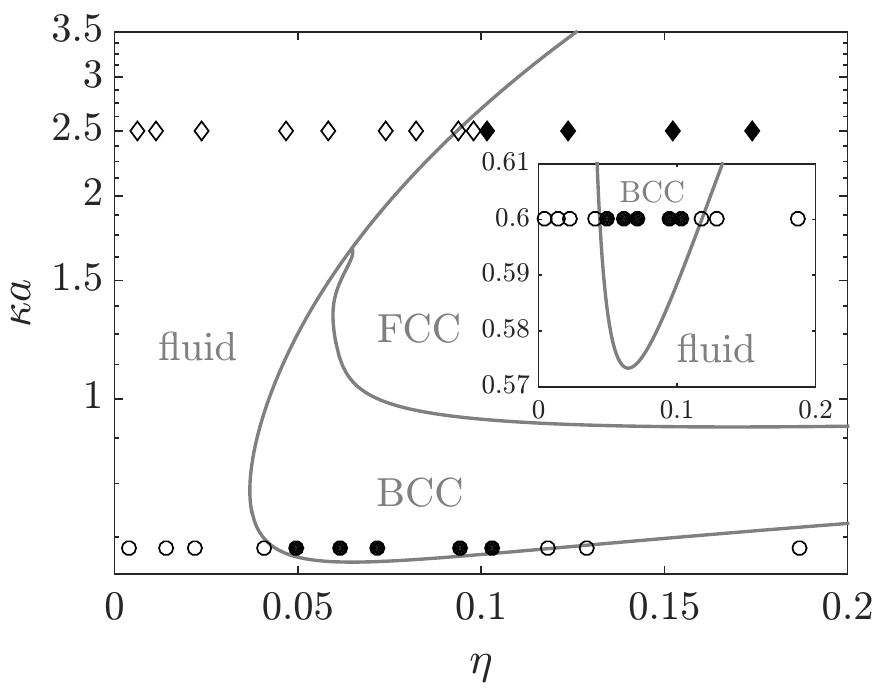}
\caption{The phase diagram calculated using the cell model in combination with Eqs. \eqref{eq:pekkaa} and \eqref{eq:pekka}, where we use the parameters $a/\lambda_B=125$ and $K_-/K_+=51$ (or equivalently $\Delta pK=pK_--pK_+=-1.7$) . The experimentally obtained state points of the fluid-FCC transition from Ref. \cite{Royall:2003} are shown as the open and filled diamonds for the fluid and FCC phase, respectively. The reentrant fluid phase of Ref. \cite{Royall:2006} are labeled by the open circles, while the BCC state points are shown as filled circles. A zoomed-in version of the phase diagram around these BCC state points are shown in the inset, to emphasize the reentrant nature of the phase transition.}
\label{fig:phasecompareexp}
\end{figure}

Further evidence for the predictive power of the present theory is provided by comparing the experimentally observed fluid-FCC phase boundary of the very same system but at a much higher salt concentration, as presented in Ref. \cite{Royall:2003}. Although no reentrant fluid was found here we can check whether $K_-/K_+=51$ that was determined for the parameters in Fig. \ref{fig:mirjam} is also able to describe this experiment. At large $\kappa$, we have $\bar{\kappa}\approx\kappa$ and for this reason we use for $\kappa a$ the value of $\bar{\kappa}a=2.5$ as was determined from the simulations in Ref. \cite{Royall:2003}. In Fig. \ref{fig:phasecompareexp} a plot is shown for the state points of the fluid (open symbols) and FCC crystal (filled symbols). For completeness we also add the state points of Fig. \ref{fig:mirjam}. We find a very good agreement of the fluid-FCC state points with the phase diagram calculated from the cell model with $K_-/K_+=51$. In Ref. \cite{Royall:2003} the radial distribution function $g(r)$ of these state points could be described by Monte-Carlo simulations with a DLVO contact potential of $\beta\epsilon_c=140$. However, we find a slightly higher contact potential from our cell model calculations,  $\beta\epsilon_c\approx200$.


\section{Discussion and conclusions}
We investigated various crystallization criteria that are based on hard spheres and Yukawa point particles. We have seen that the point-particle criteria are reliable in a large parameter regime, while the considered hard-sphere criteria should only be trusted at high salt concentrations. Moreover, we have seen that adsorption of multiple ion species on the colloidal surface can explain the occurrence of a density-induced reentrant fluid at fixed (reservoir) salt concentration. In contrast, the crystal phase is always reentrant as function of salt concentration for the charge regulation mechanisms that we investigated. We tested this result against the experiments of Refs \cite{Royall:2003, Royall:2006, Anda:2015} and found good agreement with the phase boundaries.

There are some caveats, however. In particular, we assume that the density of colloids is varied at a fixed reservoir salt concentration $\rho_s$ and hence a fixed $\kappa a$. In other words, all our calculations are based on the grand-canonical treatment of the salt. However, in experiments the ions are often treated canonically and the salt concentration is expected to change with the density of colloids. In Ref. \cite{Anda:2015}, this effect has been accounted for by defining $\tilde{\kappa}^2=8\pi\lambda_B\rho_\text{ion}$, where $\rho_\text{ion}=Z\rho+2\tilde\rho_s$ and $\tilde\rho_s$ the initial salt concentration without colloids. Here the authors interpreted the value of the inverse screening length determined from simulations of DLVO particles as $\tilde{\kappa}$, while this is $\bar{\kappa}$ within our treatment. This latter quantity does depend on colloidal density even if the ions are treated grand canonically. However, since $\bar{\kappa}\neq\tilde{\kappa}$ in general, we should perform our calculations in the canonical ensemble, by fixing the initial salt concentration and changing $\kappa$ accordingly if $\eta$ is varied. Nevertheless, we expect that this will not change the qualitative features of our result and  this may only alter the precise values of the $\epsilon_c$ and $\bar{\kappa}$ obtained from the cell model. This means that the density dependent $Z$ and $\kappa$ that was observed in Ref. \cite{Royall:2006, Anda:2015} can still be attributed to an underlying charge regulation mechanism where multiple ions are involved, regardless of this caveat. Finally, we note that there are experimental systems for which the grand-canonical treatment is justified, see for example Ref. \cite{Boon:2015}.

As an outlook we wish to state that our work also suggests that microscopic details of charging mechanisms can possibly be inferred from macroscopic measurements of phase boundaries and/or structural mesoscopic measurements of $g(r)$. The model that we presented here provides a simple of way to investigating these charge regulation effects, which are sometimes underestimated in charged colloidal suspensions.

We acknowledge fruitful discussions with C. P. Royall, M. Dijkstra and  A. van Blaaderen and financial support of a Netherlands Organisation for Scientific Research (NWO) VICI grant funded by the Dutch Ministry of Education, Culture and Science (OCW). This work is part of the D-ITP consortium, a program of the Netherlands Organisation for Scientific Research (NWO) funded by the Dutch Ministry of Education, Culture and Science (OCW).

\section*{Appendix}
In this Appendix we rationalize the constant potential behaviour that was observed in Fig. \ref{fig:charge}(b). Typically, only a small fraction of the surface sites actually contains an ion, i.e. $[\text{SP}^+]$, $[\text{SN}^-]\ll [\text{S}]$, and in this case we may approximate Eq. \eqref{eq:isotherm} as
\begin{equation}
y=Y_+\exp(-\phi_0)-Y_-\exp(\phi_0), \label{eq:simpads}
\end{equation}
where $y=Z\lambda_B/(\kappa a^2)$ and $Y_\alpha=\kappa aM/(8\pi K_\alpha a^3)$ are the dimensionless charge density and \emph{chargeability} \cite{Boon:2015} respectively, and we assume the surface to be positively charged, i.e. $Y_+ > Y_-$. The point of zero charge is given by $\bar{\phi}_0=\log(Y_+/Y_-)/2$. Expanding Eq. \eqref{eq:simpads} around this iso-electric point, gives
\begin{equation}
y=-2{\sqrt{Y_+Y_-}}[\phi_0(\eta)-\bar{\phi}_0]+\mathcal{O}\{[\phi_0(\eta)-\bar{\phi}_0]^3\},
\label{eq:taylor1}
\end{equation}
and we also find $|(\phi_0(\eta)-\bar{\phi}_0)| < |y / \sqrt{4 Y_+Y_-}|$, with $\Delta\phi_0 = (\phi_0(\eta)-\bar{\phi}_0)$. We are now interested in the maximal deviation $\Delta\phi_0=\phi_0(\eta\downarrow0)-\bar{\phi}_0$. To estimate it, we use the Gouy-Chapman relation and use it for our colloids. In this case, $y=2\sinh(\phi_0/2)$. This relation underestimates the charge at infinite dilution that comes from the cell model, but it is still a good estimate. This results in
\begin{equation}
y<  2\sinh(\bar{\phi}_0/2)<\left( Y_+/Y_-\right)^{1/4} - \left(Y_-/Y_+\right)^{1/4}, \label{eq:estimate}
\end{equation}
and thus
$|\Delta\phi_0| < \frac{1}{2} \left(Y_-^3  Y_+\right)^{-1/4} < 1/(2 Y_-) $, showing that significant chargeabilities for \emph{both} the dominant (+) as well as the competing (-) charge mechanism will lead to constant-potential like behaviour. For the light blue curves in Fig. \ref{fig:charge}(b) we have $|\Delta\phi_0|=0.07$, while the estimate in Eq. \eqref{eq:estimate} gives $|\Delta\phi_0|=0.13$. The red curve has $|\Delta\phi_0|=0.001$, while Eq. \eqref{eq:estimate} gives $|\Delta\phi_0|=0.0001$. The blue curve is not at constant potential anymore, since $|\Delta\phi_0|>1$, and this is supported by Eq. \eqref{eq:estimate} which gives $|\Delta\phi_0|=10^{2}$. Indeed the linearization in Eq. \eqref{eq:taylor1} breaks down, and the system is not constant potential.

\bibliographystyle{apsrev4-1} 
\bibliography{literature1} 

\end{document}